\documentclass[fleqn,twoside]{article}
\usepackage{espcrc2}
\usepackage{graphicx}

\title{Light meson correlation functions near the deconfining transition
  on anisotropic lattices
  \thanks{Poster presented by T.~Umeda}}

\author{
 Takashi Umeda%
  \address{Yukawa Institute for Theoretical
           Physics, Kyoto University, Kyoto 606-8502, Japan},
 Kouji Nomura%
  \address{Department of Physics, Hiroshima University, 
           Higashi-Hiroshima 739-8626, Japan \vspace{0mm}},
and
 Hideo Matsufuru$^{\rm a}$,}

\begin{document}

\begin{abstract}
We study the light hadron correlators near the deconfining transition 
by extracting the spectral function on quenched anisotropic lattices.
We adopt the method successfully applied to the charmonium systems:
the use of the smeared operators and the analysis method which is
composed of the maximum entropy method and the $\chi^2$ fitting
assuming several forms.
The numerical simulations are performed on lattices of
$a_s^{-1} \simeq 2$ GeV and $a_s/a_t=4$, with the clover quark around
the strange quark mass.
Towards applications to the plasma phase, we check the reliability
of our procedure by examining how the results for the correlators
below $T_c$ ($N_t=32$) changes under variation of input parameters
such as the smearing function.
\end{abstract}

\maketitle

\section{Introduction}

The theoretical expectation for properties of quarks and 
gluons in the deconfinement phase is strongly required for an
interpretation of experimental results in the heavy ion collision
experiments.
Contrary to a naive quark gluon plasma (QGP) picture such as
the gas of weakly interacting quarks and gluons, several model
calculations inspired by QCD indicate nonperturbative features of QGP.
For example, an NJL model calculation suggests that the hadronic
excitations survive as soft modes even above the critical temperature
\cite{hatsuda85}. 

The lattice QCD can in principle solve the problem
in a model independent way, since the dynamical properties of
excitation modes can be extracted from the Matsubara Green functions
\cite{abrikovov59}.
Although in practice such studies are quite nontrivial even at
qualitative level,
recent technical development such as anisotropic lattices and
computational progress have enabled us to study
dynamical properties of hadrons at finite temperature
\cite{forcrand01,umeda01}.
In addition, several procedures such as the maximum entropy method 
(MEM) \cite{nakahara99} have been developed to analyze
the spectral function with which we can directly probe the excitation 
modes of the plasma phase.
These procedures have been applied to the charmonium systems
and suggest that hadronic excitations survive even above the
critical temperature \cite{umeda03,petreczky03,asakawa03}.

In this work we apply the method which was adopted
in Ref.~\cite{umeda03} to systems with light quarks. 
In this report we focus on a check of reliability for applied
analysis methods at finite temperature, as a preparation
for future applications to the deconfinement phase.

\section{Our approach}

Here we briefly explain our analysis procedure and its criteria
for a reliability, which were in detail described
in Ref.~\cite{umeda03}.

In order to extract reliable information on the spectral function,
we use MEM and $\chi^2$ fit method in a complementary manner.
After a rough estimation of spectral function by MEM,
we evaluate properties of the mode such as mass and width more 
quantitatively by the $\chi^2$ fit assuming several fitting forms
based on MEM results.
As more sophisticated form of the $\chi^2$ fit we adopt the constrained
curve fitting (CCF) \cite{repage02} whose prior knowledge is also
estimated with the MEM results.

To verify the reliability of our analysis,
we require the following criteria for the extraction methods: 
(1) The stability of the spectral function against variations of
 input parameters or model functions. 
(2) The stability of the result for the correlators at $T=0$ under
restriction of the degrees of freedom.
For the latter, since the $T=0$ correlators are not ready at present,
we instead analyze those below $T_c$ which should produce similar
results as at $T=0$.

In the previous work for the charmonium, in order to satisfy the
criteria we employ the smearing operators 
which enhance the low frequency part of the meson correlators.
However the smeared operators may produce artificial peaks
in the spectral function \cite{wetzorke02}.
In order to distinguish an artificial peak from the genuine physical
one we apply several smearing functions and examine
the stability of the results.

\section{Numerical results}

We use quenched lattices of the sizes $20^3\times N_t$, where
$N_t=160$ and 32 which roughly correspond to $T\simeq 0$ and
0.9$T_c$, respectively.
The zero temperature lattice, and the setup of lattice parameters
are the same as those used in Ref.~\cite{matsufuru01}.
The gauge configurations are generated with the Wilson plaquette
action at $\beta=6.10$ with the renormalized anisotropy 
$\xi=a_{\sigma}/a_t = 4$.
The spatial cutoff set by the hadronic radius $r_0$ is
$a^{-1}_{\sigma}=2.030(13)$ GeV.
$N_t=28$ is close to the phase transition.
The quark action is the $O(a)$ improved Wilson action \cite{matsufuru01},
with the hopping parameter roughly corresponding to the strange quark 
mass.
We have 500 configurations at $T\!=\!0$ and 1000 configurations at $T>0$.

We measure the point correlators at $T=0$ and the smeared correlators
at $T\simeq 0.9 T_c$.
For the latter, the operators are spatially extended by Gaussian
functions $\phi(\vec{x})\propto \exp{(a|\vec{x}|^2)}$
with $a=0.16$ and 0.037 which are referred as ``smeared'' and 
``half-smeared'', respectively.

\begin{figure}[tb]
\vspace{3mm} \hspace{0mm}
\includegraphics[width=68mm]{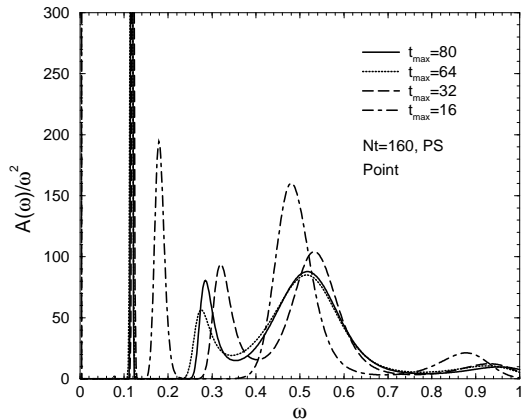}
\vspace{-11mm}
\caption{$t_{max}$ dependence of the spectral function
for the point correlator in PS channel.}
\label{fig1}
\vspace{-7mm}
\end{figure}

\begin{figure}[tb]
\vspace{3mm} \hspace{2mm}
\includegraphics[width=66mm]{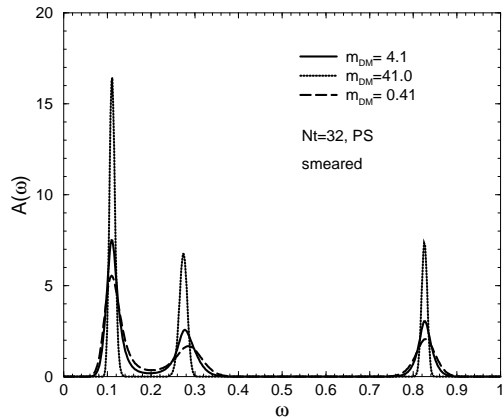}
\vspace{-11mm}
\caption{The spectral function for the smeared PS correlator
at $T\simeq 0.9T_c$.
The default model is given by $m(\omega)=m_{DM}\omega^2$.}
\label{fig2}
\vspace{-7mm}
\end{figure}

First we present a result for the point operators.
Using the correlators at $T=0$ we test the applicability to 
finite temperatures by varying the number of degrees of freedom.
Figure~\ref{fig1} shows the $t_{max}$ dependence of the MEM results,
where the fit ranges are restricted to $t=1 \sim t_{max}$.
The result at $t_{max}=16$ can not reproduce even the position of the
lowest peak, while $t_{max}=16$ corresponds to the similar situation
we encounter at $T\sim 0.9 T_c$.
This is one of the reasons why we give up to use the point operators
to explore finite temperatures.

The MEM result with the smeared PS correlator at $T\simeq 0.9T_c$
are presented in Fig.~\ref{fig2}. 
The lowest and next-lowest peaks are located at almost the same positions
as at $T=0$ which is expected to have a similar result to that
at $T\simeq 0.9 T_c$. 
This is encouraging for applications of the MEM to the systems
at $T>0$.
Since the results are sensitive to the default model function
(as input parameters), it is difficult to obtain quantitative
results only with MEM.

\begin{figure}[tb]
\vspace{3mm} \hspace{0mm}\noindent
\hspace*{3mm}
\includegraphics[width=65mm]{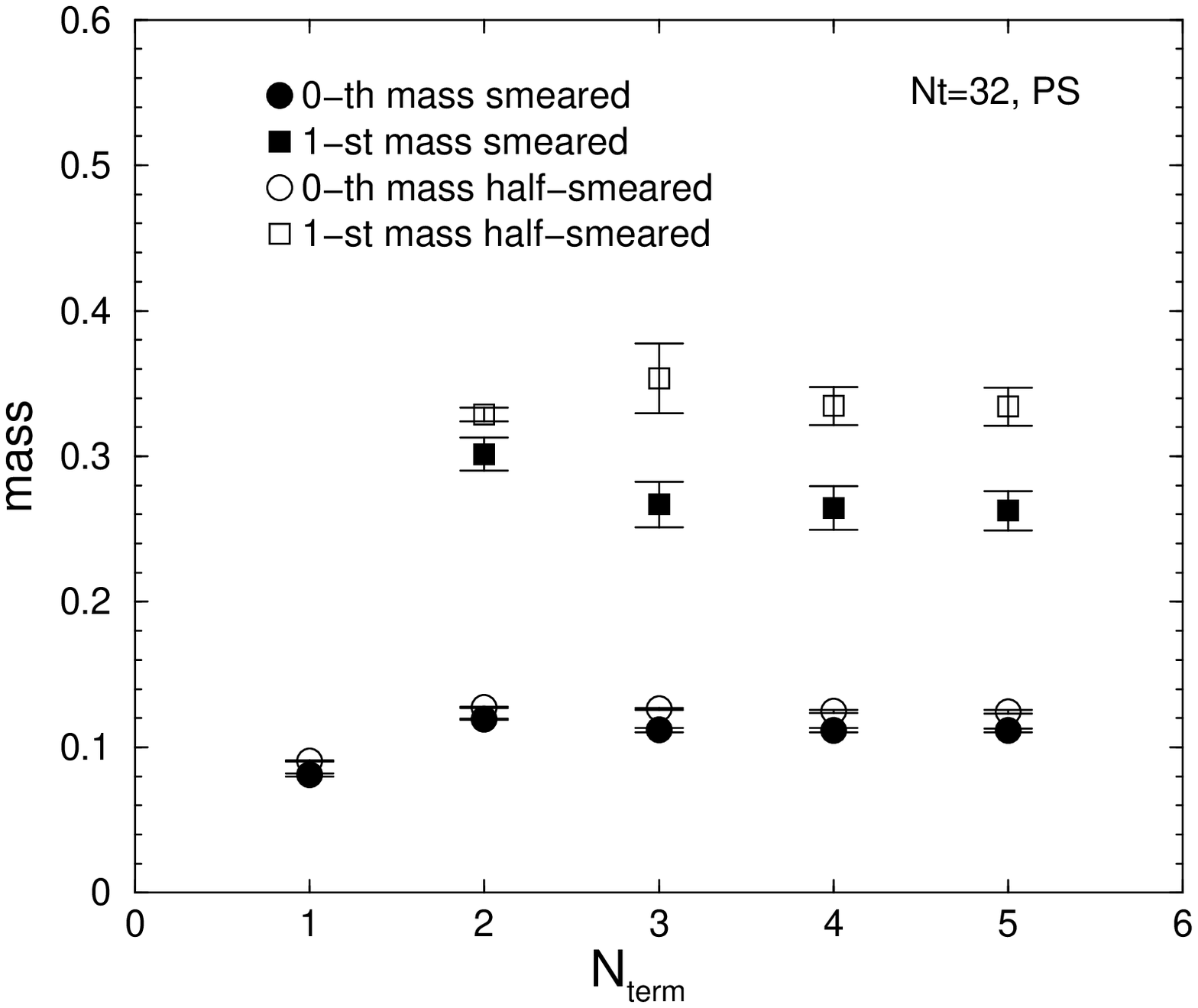}\\[1mm]
\hspace*{3mm}
\includegraphics[width=65mm]{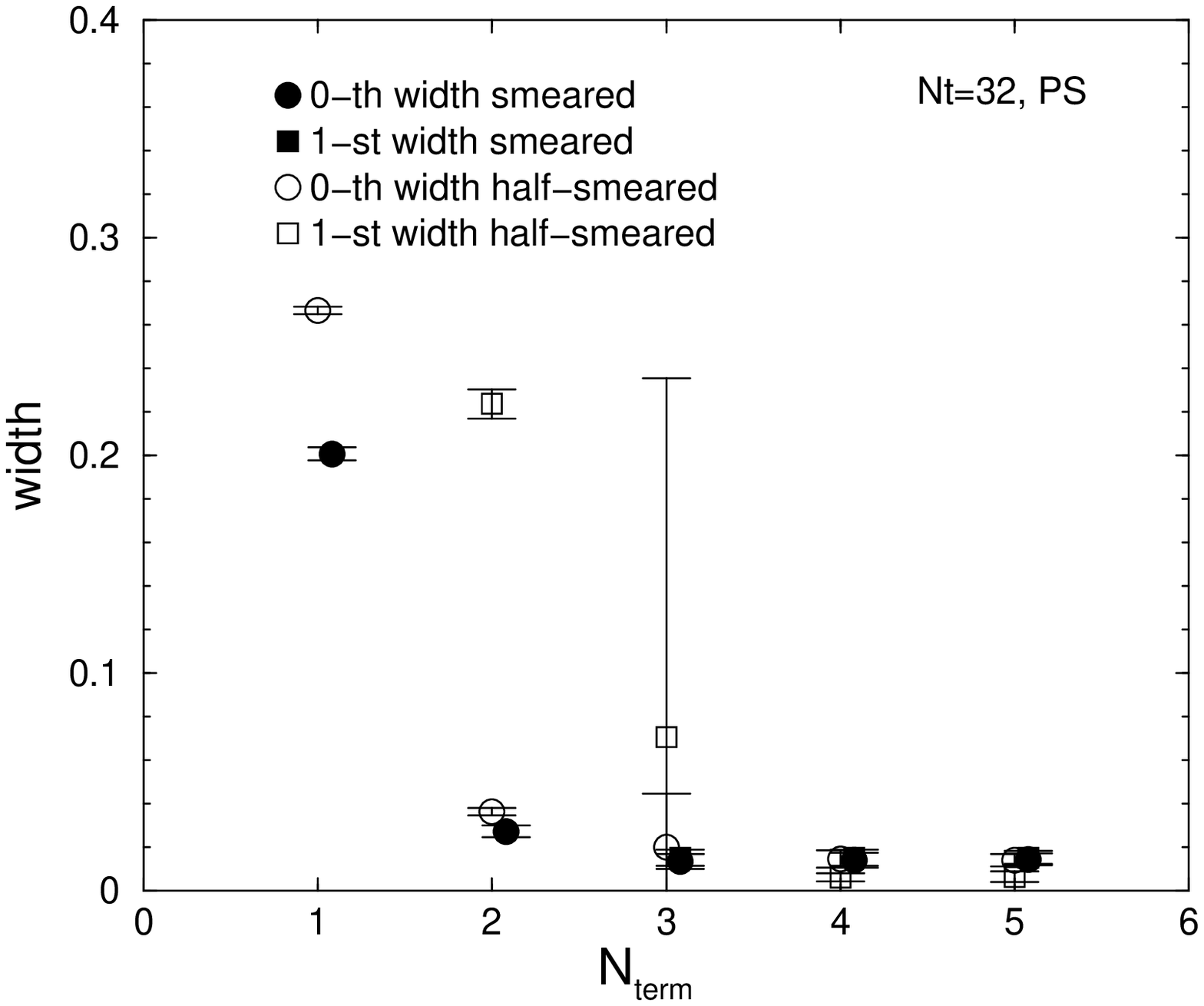}
\vspace{-10mm}
\caption{$N_{term}$ dependences of mass and width by CCF
for the smeared and half-smeared PS correlators.}
\label{fig3}
\vspace{-5mm}
\end{figure}

Based on the MEM results in Fig.~\ref{fig2}, we assume multi-peak
functions with the Breit-Wigner type forms,
\begin{eqnarray}
 A(\omega)=\sum_{i=1}^{N_{term}}\frac{\omega^2m_i\gamma_i R_i}
{(\omega^2-m_i^2)^2+m_i^2\gamma_i^2},
\end{eqnarray} 
as fitting functions and estimate the prior knowledge of the CCF.
In Fig.~\ref{fig3} the fit results for masses and widths of the lowest
and next-lowest peaks are displayed, where the results for smeared
and half-smeared correlators are plotted with different symbols.
Although with a sufficiently large $N_{term}$ $(\ge 4)$ the results
are stable, the differences between the values for the smeared and
half-smeared correlators are larger than the statistical errors.
Furthermore the results are rather sensitive to the prior
knowledge.
These results indicate that we need further investigations
of CCF to obtain quantitative results with
keeping systematic uncertainties under control at finite temperature.

Following our strategy, we have to perform these reliability checks
with correlators at $T=0$ and to examine the $t_{max}$ dependence
of the systematic uncertainties.
We also need further examination of the smearing operators, 
because the suppression of high frequency part of the correlators
may not be sufficient with the present smearing operators.
This might be a reason that CCF does not work stably at
$T\simeq 0.9 T_c$.
These problems must be clarified for reliable analysis of
the hadronic correlators at finite temperature, in particular
in the plasma phase.

This study was performed on the SR8000 at KEK (High Energy
Accelerator Research Organization) and 
the SX-5 at RCNP (Research Center for Nuclear Physics).  
T.U and H.M are supported by the JSPS Research Fellowships
for Young Scientists.

\end{document}